\documentclass[showpacs,preprintnumbers,superscriptaddress]{revtex4}
\usepackage{CJK}
\usepackage{amsmath,amssymb,graphicx,bm}
\begin{document}

\title{Effects of quantum fluctuations of metric on the universe}

\author{Rongjia Yang \footnote{Corresponding author}}
\email{yangrongjia@tsinghua.org.cn}
\affiliation{College of Physical Science and Technology, Hebei University, Baoding 071002, China}
\affiliation{Hebei Key Lab of Optic-Electronic Information and Materials}
\affiliation{State Key Laboratory of Theoretical Physics, Institute of Theoretical Physics, Chinese Academy of Sciences,
Beijing 100190, China}

\begin{abstract}
We consider a model of modified gravity from the nonperturbative quantization of a metric. We obtain the modified gravitational field equations and the modified conservational equations. We apply it to the FLRW spacetime and find that due to the quantum fluctuations a bounce universe can be obtained and a decelerated expansion can also possibly be obtained in a dark energy dominated epoch. We also discuss the effects of quantum fluctuations on inflation parameters (such as slow-roll parameters, spectral index, and the spectrum of the primordial curvature perturbation) and find values of parameters in the comparing the predictions of inflation can also work to drive the current epoch of acceleration. We obtain the constraints on the parameter of the theory from the observation of the big bang nucleosynthesis.
\end{abstract}


\maketitle

\section{Introduction}
A great number of astronomical observations have confirmed that the Universe is experiencing an accelerated expansion. In the framework of general relativity an unknown energy component, dubbed as dark energy, is usually introduced to explain this phenomenon. The simplest candidate of dark energy is the vacuum energy with a constant equation of state (EoS) parameter $w=-1$. This model is consistent with most of the current astronomical observations, but it suffers from the cosmological constant problem \citep{Carroll:2000fy} and age problem \citep{Yang:2009ae} as well. Therefore it is natural to consider more complicated cases.
The most popular attempt is to propose modifications of the Einstein-Hilbert Lagrangian by adopting different functions of the Ricci scalar, known as $f(R)$ theories, which have been studied extensively \citep{Capozziello:2011et,DeFelice:2010aj,Sotiriou:2008rp,Silvestri:2009hh,Nojiri:2010wj}. However, the fourth order field equations in $f(R)$ theories make it is hard to analyze. Analogous to $f(R)$ theories, a new scenario based on the modification of the teleparallel gravity, called $f(T)$ theory, was proposed to explain the accelerated expansion of the Universe \citep{Bengochea:2008gz,Yang:2010hw}. Recently, it has been shown that one can obtain modified gravity from Heisenberg's nonperturbative quantization \citep{Dzhunushaliev:2013nea,Dzhunushaliev:2015mva,Dzhunushaliev:2015hoa}. According to this technique, the classical fields appearing in the corresponding field equation are replaced by operators of the fields. Think of general relativity, we have the operator Einstein equations
\begin{eqnarray}
\label{eins}
\hat{G}_{\mu\nu}\equiv \hat{R}_{\mu\nu}-\frac{1}{2}\hat{g}_{\mu\nu}\hat{R}=k^2 \hat{T}_{\mu\nu},
\end{eqnarray}
where $k^2=8\pi G$ and we take $c=1$. All geometric operators, such as $\hat{\Gamma}^a_{~bc}$, $\hat{R}^a_{~bcd}$, and $\hat{R}_{ab}$, are defined in the same way as in the classical case by replacing the classical quantities with the corresponding operators \citep{Dzhunushaliev:2013nea}. Heisenberg's technique offers one possibility to solve this operator equation by average it over all possible products of the metric operator $\hat{g}(x_1),...,\hat{g}(x_n)$ which can be written as an infinite set of equations for all Green's functions:

\begin{eqnarray}
\label{green}
\langle Q|\hat{g}(x_1)\cdot\hat{G}_{\mu\nu}|Q \rangle &=&k^2 \langle Q|\hat{g}(x_1)\cdot\hat{T}_{\mu\nu}|Q\rangle,\\
...&=&...,\\
\langle Q|\hat{g}(x_1)\cdot...\cdot\hat{g}(x_n)\cdot\hat{G}_{\mu\nu}|Q \rangle &=&k^2 \langle Q|\hat{g}(x_1)\cdot...\cdot\hat{g}(x_n)\cdot\hat{T}_{\mu\nu}|Q\rangle,
\end{eqnarray}
where $|Q \rangle$ is a quantum state \citep{Dzhunushaliev:2013nea,Dzhunushaliev:2015mva}. This set of equations cannot be solved analytically. Some proximate methods to solve them were discussed in Refs \citep{Dzhunushaliev:2013nea,Dzhunushaliev:2012vb}, for example, one can decompose the metric operator into a sum of an average metric $g_{\mu\nu}$ and a fluctuating part $\widehat{\delta g}_{\mu\nu}$ \citep{Dzhunushaliev:2013nea}.
\begin{eqnarray}
\label{met}
\hat{g}_{\mu\nu}=g_{\mu\nu}+\widehat{\delta g}_{\mu\nu}.
\end{eqnarray}
Assuming $\langle\delta \hat{g}_{\mu\nu}\rangle\neq 0$ and ignoring high order fluctuations, we expand the Einstein-Hilbert Lagrangian $L_{\hat{g}}=\frac{1}{2k^2}\sqrt{-\hat{g}} \hat{R}$ in the following manner
\begin{eqnarray}
\label{lag0}
L_{\hat{g}}=L_{\hat{g}}(g+\delta\hat{g})\approx L_{g}(g)+\frac{\delta L_{g}}{\delta g^{\mu\nu}}\widehat{\delta g}^{\mu\nu}.
\end{eqnarray}
The expectation value of $\frac{\delta L_{g}}{\delta g^{\mu\nu}}\widehat{\delta g}^{\mu\nu}$ can be represented as $\left\langle \frac{\delta L_{g}}{\delta g^{\mu\nu}}\widehat{\delta g}^{\mu\nu}\right\rangle=\frac{\delta L_{g}}{\delta g^{\mu\nu}}\langle \widehat{\delta g}^{\mu\nu}\rangle=\sqrt{-g}G_{\mu\nu}\langle \widehat{\delta g}^{\mu\nu}\rangle$ \cite{Dzhunushaliev:2013nea}. Then the expectation value of Lagrangian (\ref{lag0}) has the form 
\begin{eqnarray}
\label{lag00}
\langle L_{\hat{g}}\rangle\approx \frac{1}{2k^2}\sqrt{-g} \big[R+G_{\mu\nu}\langle\delta \hat{g}^{\mu\nu}\rangle \big].
\end{eqnarray}
Similarly, we can expand the quantum Lagrange density $L^{\hat{g}}_{\rm m}$ as follows
\begin{eqnarray}
\label{lagm}
L^{\hat{g}}_{\rm m}(g+\delta\hat{g})\approx \sqrt{-g}L_{\rm m}(g)+\frac{\delta \sqrt{-g}L_{\rm m}}{\delta g^{\mu\nu}}\widehat{\delta g}^{\mu\nu}.
\end{eqnarray}
With the aforementioned assumptions, the expectation value of the quantum Lagrange density (\ref{lagm}) takes the form \citep{Dzhunushaliev:2013nea}
\begin{eqnarray}
\label{lagm0}
\langle L^{\hat{g}}_{\rm m}(g+\delta\hat{g}) \rangle \approx \sqrt{-g}\big[L_{\rm m}+\frac{1}{2}T_{\mu\nu}\langle\delta \hat{g}^{\mu\nu}\rangle\big].
\end{eqnarray}
Therefore the modified Lagrangian density can be written as (for details, see Ref. \citep{Dzhunushaliev:2013nea,Dzhunushaliev:2015mva})
\begin{eqnarray}
\label{lag}
L=\frac{1}{2k^2}\sqrt{-g} \big[R+G_{\mu\nu}\langle\delta \hat{g}^{\mu\nu}\rangle \big]+\sqrt{-g}\big[L_{\rm m}+\frac{1}{2}T_{\mu\nu}\langle\delta \hat{g}^{\mu\nu}\rangle\big].
\end{eqnarray}

In this paper, we will consider the simplest case: $\langle\delta g^{\mu\nu}\rangle=\alpha g^{\mu\nu}$ and derive the equation of motion. Then we will investigate the effects of quantum fluctuations of metric on the evolution of the universe. Obviously, if $|\alpha|\ll 1$, the modified Lagrangian density (\ref{lag}) approximates to the Hilbert-Einstein Lagrangian density, namely the model of modified gravity proposed here approximates to general relativity. In order to adequately analyze the effects of quantum fluctuations on the universe, or in other words, to take seriously the way to modify gravity from the nonperturbative quantization of a metric \citep{Dzhunushaliev:2013nea}, we will consider all possible values the parameter $\alpha$ can take.

The paper is outlined as follows. In next section, we will present the modified gravitational field equations and the modified conservational equations. In Sec. III, effects of quantum fluctuations on the universe are discessed. Finally, we will briefly summarize and discuss our results in section IV.

\section{Modified gravitational field equations}
For Lagrangian density (\ref{lag}), since $\langle\delta \hat{g}_{\mu\nu}\rangle$ has the same symmetry as the metric $g_{\mu\nu}$, the simplest case is $\langle\delta \hat{g}^{\mu\nu}\rangle=\alpha g^{\mu\nu}$, which was suggested in \citep{Dzhunushaliev:2013nea}. Therefore Lagrangian density (\ref{lag}) takes the form
\begin{eqnarray}
\label{act}
L=L_{\rm mg}+L_{\rm mm}=\frac{1}{2k^2}\sqrt{-g}(1-\alpha)R+\sqrt{-g}[L_{\rm m}-\frac{1}{2}\alpha T],
\end{eqnarray}
where $T=g_{\mu\nu}T^{\mu\nu}$ with $T_{\mu\nu}=-2\delta(\sqrt{-g}L_{\rm m})/(\sqrt{-g}\delta g^{\mu\nu})$. Lagrangian density (\ref{act}) may can be rewritten as a special type of $f(R,T)$ gravity phenomenologically proposed in \citep{Harko:2011kv}, but here we obtain it basing on theoretical considerations and for the first time discuss it in detail. In general, comparing with the classical part of the metric, quantum fluctuations are small ($|\alpha|<1$), such as in radiation, matter, or dark energy dominated era; however, it is possible that quantum fluctuations are large ($|\alpha|>0$) when the system we consider closes to the Planck scale, such as in the very early epoch. Varying Lagrangian density (\ref{lag}) with respect to $g^{\mu\nu}$ and assuming $\delta g_{\mu\nu}=0$ on the boundary, we obtain
\begin{eqnarray}
\label{var1}
\delta L_{\rm mg}&=&\frac{1}{2k^2}(1-\alpha)\left(R_{\mu\nu}-\frac{1}{2}g_{\mu\nu}R \right)\delta g^{\mu\nu},\\
\label{var2}
\delta(\sqrt{-g}L_{\rm m})&=&-\frac{1}{2}T_{\mu\nu}\sqrt{-g}\delta g^{\mu\nu},
\end{eqnarray}
and
\begin{eqnarray}
\label{var3}
\delta(\sqrt{-g}T)=T\delta\sqrt{-g}+\sqrt{-g}\delta T=-\frac{1}{2}g_{\mu\nu}T\sqrt{-g}\delta g^{\mu\nu}+\sqrt{-g}(T_{\mu\nu}+\theta_{\mu\nu})\delta g^{\mu\nu},
\end{eqnarray}
where $\delta T/\delta g^{\mu\nu}=T_{\mu\nu}+\theta_{\mu\nu}$ with $\theta_{\mu\nu}=g^{\alpha\beta}\delta T_{\alpha\beta}/\delta g^{\mu\nu}$. Assuming that the Lagrangian density $L_{\rm m}$ of matter depends only on
the metric tensor component $g_{\mu\nu}$ , not on its derivative, one can easily get $T_{\mu\nu}=g_{\mu\nu}L_{\rm m}-2\partial L_{\rm m}/\partial g^{\mu\nu}$ and
\begin{eqnarray}
\frac{\delta T_{\alpha\beta}}{\delta g^{\mu\nu}}=-g_{\alpha\lambda}g_{\beta\sigma}\delta^{\lambda\sigma}_{\mu\nu}L_{\rm m}+\frac{1}{2}g_{\mu\nu}g_{\alpha\beta}L_{\rm m}-\frac{1}{2}T_{\mu\nu}g_{\alpha\beta}-2\frac{\partial^2L_{\rm m}}{\partial g^{\alpha\beta}\partial g^{\mu\nu}},
\end{eqnarray}
where $\delta^{\lambda\sigma}_{\mu\nu}=\delta g^{\lambda\sigma}/\delta g^{\mu\nu}$. Then we have
\begin{eqnarray}
\theta_{\mu\nu}=g_{\mu\nu}L_{\rm m}-2T_{\mu\nu}-2g^{\alpha\beta}\frac{\partial^2L_{\rm m}}{\partial g^{\alpha\beta}\partial g^{\mu\nu}}.
\end{eqnarray}
For different matter, $\theta_{\mu\nu}$ takes different form. From equations (\ref{var1}), (\ref{var2}), and (\ref{var3}), we obtain the gravitational field equations
\begin{eqnarray}
\label{mot}
R_{\mu\nu}-\frac{1}{2}g_{\mu\nu}R=\frac{2k^2}{1-\alpha}\left[\frac{1}{2}(1+\alpha)T_{\mu\nu}-\frac{1}{4}\alpha g_{\mu\nu}T+\frac{1}{2}\alpha\theta_{\mu\nu}\right].
\end{eqnarray}
We can seen that the gravitational constant $G$ and the energy-momentum tensor are modified due to the quantum fluctuation of the metric. Because $R=-k^2[T+\alpha\theta/(1-\alpha)]$, the gravitational field equations (\ref{mot}) can be rewritten as
\begin{eqnarray}
\label{mot1}
R_{\mu\nu}=\frac{2k^2}{1-\alpha}\left[\frac{1}{2}(1+\alpha)T_{\mu\nu}-\frac{1}{4} g_{\mu\nu}T+\frac{1}{2}\alpha\theta_{\mu\nu}-\frac{1}{4}\alpha g_{\mu\nu}\theta\right].
\end{eqnarray}
Taking into account the covariant divergence of Einstein tensor $\nabla_{\nu}G^{\mu\nu}=0$, we get
for the divergence of the stress-energy tensor $T_{\mu\nu}$ the equation
\begin{eqnarray}
\label{conser}
\nabla^{\nu}T_{\mu\nu}=\frac{1}{1+\alpha}\left[\frac{1}{2}\alpha \nabla_{\mu}T-\alpha\nabla^{\nu}\theta_{\mu\nu}\right].
\end{eqnarray}
In other words, if we take into account of quantum fluctuations, the stress-energy tensor are not conserved quantities any more. Discussions on the nonconservation of stress-energy tensor can be found in \citep{Bertolami:2008ab,Harko:2008qz,Bisabr:2012tg,Minazzoli:2013bva}. Obviously, for small $\alpha$, the effects of quantum fluctuations are weak, Lagrangian density (\ref{lag}) approximates to the Hilbert-Einstein Lagrangian density. In order to adequately investigate the effects of quantum fluctuations of metric on the universe, see, for example, for a $\alpha\sim 1$, the effects of quantum fluctuations can even approximatively counteract the gravity, it is worth considering all possible values the parameter $\alpha$ can take. We will discuss this topic in detail in the following sections.

\section{Cosmological applications}
In this section, we apply the gravitational field equations (\ref{mot}) to a homogeneous and isotropic Friedmann-Lema\^{i}tre-Robertson-Walker (FLRW) universe with scalar factor $a$,
\begin{align}
ds^2=dt^2-a^2(t)\left[\frac{dr^2}{1+Kr^2}+r^2(d\theta^2+\sin^2\theta d\phi^2)\right],
\end{align}
where the spatial curvature constant $K=+1$, 0, and $-1$ correspond to a closed, flat, and open universe, respectively. In the present study, we consider a perfect fluid specified by the stress energy tensor
\begin{eqnarray}
T_{\mu\nu}=(\rho+p)u_{\mu}u_{\nu}-g_{\mu\nu}p.
\end{eqnarray}
There are different choices for the Lagrangian density of the perfect fluid and all of them leads to the same field equations and the stress-energy tensor in the context of general relativity \citep{PhysRevD.2.2762, 0264-9381-10-8-017}. Two Lagrangian densities, $L_{\rm m}=-p$ and $L_{\rm m}=\rho$, have been widely used in the literatures \citep{0264-9381-25-20-205002, PhysRevD.80.124040, PhysRevD.75.104016, PhysRevD.81.104046, PhysRevD.85.024012, PhysRevD.83.044010}. The matter Lagrangian density as an arbitrary function of the energy density of the matter $\rho$ only has been discussed in \citep{Harko:2008qz}. Here the matter Lagrangian density $L_{\rm m}$ enters explicitly the field equations (\ref{mot}) and all results strongly depend on the choice of $L_{\rm m}$. Following Ref. \citep{Harko:2011kv}, we take the Lagrangian for matter as $L_{\rm m}=-p$. The fluid 4-velocity satisfies the condition $u^{\mu}u_{\mu}=1$. Then for the variation of the stress-energy of a perfect
fluid, we get the expression
\begin{eqnarray}
\theta_{\mu\nu}=-2T_{\mu\nu}-g_{\mu\nu}p.
\end{eqnarray}
With $T=\rho-3p$ and $\theta\equiv \theta^\mu_\mu=-2T-4p=-2(\rho-p)$, the $\mu\nu=00$ component of the gravitational field equations (\ref{mot1}) is
\begin{eqnarray}
\label{motf}
\frac{\ddot{a}}{a}=-\frac{1}{3}\frac{k^2}{2(1-\alpha)}\big[\rho+(3-4\alpha)p\big],
\end{eqnarray}
and equations $\mu\nu=00$ and $\mu\nu=11$ together gives
\begin{eqnarray}
\label{motf1}
H^2+\frac{K}{a^2}=\frac{1}{3}\frac{k^2}{2(1-\alpha)}\left[(2-3\alpha)\rho+\alpha p\right].
\end{eqnarray}
To investigate the effects of quantum fluctuations of metric on the universe, for simplicity while without losing generality here we consider a spatial flat FLRW spacetime. Equations (\ref{motf1}) and (\ref{motf}) are rewritten as
\begin{eqnarray}
\label{motfd1}
H^2&=&\frac{k^2}{3}\left[\frac{2-3\alpha}{2-2\alpha}\rho+\frac{\alpha}{2-2\alpha} p\right]=\frac{k^2}{3}\left[\frac{2-(3-w)\alpha}{2-2\alpha} \right]\rho,\\
\label{motfd2}
\frac{\ddot{a}}{a}&=&-\frac{k^2}{3}\frac{1}{2(1-\alpha)}\big[\rho+(3-4\alpha)p\big]=-\frac{k^2}{6}\left[\frac{1+(3-4\alpha)w}{1-\alpha}\right]\rho,
\end{eqnarray}
For three cases: (1) $w=0$ and $\alpha>1$, (2) $0<w<1$ and $1<\alpha<\frac{3}{4}+\frac{1}{4w}$, (3) $w<0$ and $\alpha<\frac{3}{4}+\frac{1}{4w}$ or $\alpha>1$, we have $\ddot{a}>0$, an accelerated expansion can be obtained. For $\alpha=\frac{3}{4}+\frac{1}{4w}$, a constant-speed expansion is obtained. For other cases, decelerated expansions are obtained.

Contracting equations (\ref{conser}) with $u^\mu$, we obtain
\begin{eqnarray}
u^\mu\nabla^\nu T_{\mu\nu}&=&\dot{\rho}+3H(\rho+p),\\
u^\mu\nabla_\mu T&=&\dot{\rho}-3\dot{p},\\
u^\mu\nabla^\nu \theta_{\mu\nu}&=&-2[\dot{\rho}+3H(\rho+p)]-\dot{p}.
\end{eqnarray}
The modified conservation equation (\ref{conser}) takes the form
\begin{eqnarray}
\label{cons1}
(1-\frac{3}{2}\alpha+\frac{1}{2}w\alpha)\dot{\rho}+\frac{1}{2}\alpha\rho \dot{w}&=&3(\alpha-1)H\rho(1+w).
\end{eqnarray}
For $\alpha=0$, equations (\ref{motfd1}), (\ref{motfd2}), and (\ref{cons1}) reduce to the standard forms.

Now we investigate the effects of the quantum fluctuation of metric on the very early universe, inflation, the radiation-dominated, the dust-dominated, or the dark energy dominated epochs.

We first consider the effects of the quantum fluctuations on the very early universe. In this era, it is possible that the quantum fluctuations may be large. From equations (\ref{motfd1}) and (\ref{motfd2}), we find that $H=0$ and $\dot{H}>0$ will be satisfied for $\alpha=2/(3-w)$ and $-1<w<0$ (implying $1/2<\alpha<2/3$), namely a bounce universe can be obtained, which can not be realized in general relativity with the same matter.

Secondly, we discuss the effects of the quantum fluctuations on inflation. We consider a scalar field $\phi$, which has the energy density, $\rho=\frac{1}{2}\dot{\phi}^2+V(\phi)$, and the pressure, $\rho=\frac{1}{2}\dot{\phi}^2-V(\phi)$. The conditions for slow-roll inflation are: $\dot{\phi}^2\ll V(\phi)$ and $\ddot{\phi}\ll dV(\phi)/d\phi$. Therefore, equations (\ref{motfd1}) and (\ref{cons1}) take the form, respectively
\begin{eqnarray}
\label{inf}
H^2=\frac{k^2}{3}\frac{1-2\alpha}{1-\alpha}V(\phi),\\
\label{inf1}
3H\dot{\phi}=-\frac{1-2\alpha}{1-\alpha}\frac{dV(\phi)}{d\phi}.
\end{eqnarray}
Comparing with the standard equations of inflation, an additional factor $\frac{1-2\alpha}{1-\alpha}$ appears due to the quantum fluctuations of the metric, which may have effects on the process of the inflation. Condition $\frac{1-2\alpha}{1-\alpha}>0$ gives: $\alpha>1$ or $\alpha<1/2$. The two slow-roll parameters are modified as, respectively
\begin{eqnarray}
\label{param}
\epsilon_{\rm M}=\left(\frac{1-2\alpha}{1-\alpha}\right)\epsilon,\\
\label{param1}
\eta_{\rm M}=\left(\frac{1-\alpha}{1-2\alpha}\right)\eta,
\end{eqnarray}
where $\epsilon\equiv \frac{1}{24\pi G} \left(\frac{V'}{V}\right)^2$ and $\eta\equiv \frac{1}{8\pi G} \frac{V''}{V}$ are the standard slow-roll parameters. The spectral index and the spectrum of the primordial curvature perturbation are modified as, respectively
\begin{eqnarray}
\label{n}
n_{\rm sM}=1-6\epsilon_{\rm M}+2\eta_{\rm M}&=&1-6\left(\frac{1-2\alpha}{1-\alpha}\right)\epsilon+2\left(\frac{1-\alpha}{1-2\alpha}\right)\eta,\\
P_{\mathcal{R\rm{M}}}&=&\left(\frac{1-\alpha}{1-2\alpha}\right)P_\mathcal{R},
\end{eqnarray}
where $P_\mathcal{R}$ is the standard spectrum. Due to the quantum fluctuations of the metric, modified $\eta_{\rm M}$ and $P_{\mathcal{R\rm{M}}}$ are larger than the standard ones for $0<\alpha<1/2$ or $\alpha>1$ and are smaller than the standard ones for $\alpha<0$, while $\epsilon_{\rm M}$ changes on the contrary.

According to the Planck 2015 results, the spectral index of curvature perturbations is constrained as $n_{\rm s}=1-6\epsilon+2\eta=0.968\pm 0.006$ \cite{Ade:2015lrj}. Here we take $\epsilon=0.01$ and $\eta=0.014$ for quantitative analysis. From equation (\ref{n}) and taking $\alpha=0.01$, we get $n_{\rm sM}=0.965$ and $P_{\mathcal{R\rm{M}}}=1.01 P_\mathcal{R}$; taking $\alpha=-0.01$, we have $n_{\rm sM}=0.963$ and $P_{\mathcal{R\rm{M}}}=0.99 P_\mathcal{R}$. All these results are consistent with the Planck observations.

For a constant EoS, equation (\ref{cons1}) takes the form
\begin{eqnarray}
\label{cons2}
(1-\frac{3}{2}\alpha+\frac{1}{2}w\alpha)\dot{\rho}=3(\alpha-1)(1+w)H\rho,
\end{eqnarray}
which can be integrated as
\begin{eqnarray}
\label{integ}
\rho=\rho_0a^{\frac{6(\alpha-1)(1+w)}{2-3\alpha+w\alpha}}.
\end{eqnarray}
For a radiation-dominated epoch, $w=1/3$, from equation (\ref{integ}) we get
\begin{eqnarray}
\label{er}
\rho=\rho_0a^{\frac{12(\alpha-1)}{3-4\alpha}}.
\end{eqnarray}
It reduces to the standard evolution of radiation: $\rho=\rho_0a^{-4}$ for $\alpha=0$. If $1<\alpha<3/2$, namely, the quantum fluctuations are large, the expansion of the universe will be accelerated even in a radiation-dominated period. However, it can be expected that the quantum fluctuations in the this era should be very small. Taking $\alpha=0.01$, it is easy to get $\rho=\rho_0a^{-4-\frac{1}{74}}\simeq \rho_0a^{-4}$.
Taking $\alpha=-0.01$, we have $\rho=\rho_0a^{-4+\frac{1}{76}}\simeq \rho_0a^{-4}$.

From equation (\ref{motfd1}), the quantum fluctuations of metric could lead to a non-standard, early universe expansion rate $H'$, whose ratio to the standard rate $H$ is parameterized by an expansion rate factor
$S\equiv H'/H$. The modification of expansion might also arise due to additional light particle such as neutrinos which would make the ratio to be $H'/H=[1+7(N_{\nu}-3)/43]^{1/2}$. Here we are interested in the modifications coming for the quantum fluctuations of metric and take $N_{\nu}=3$. As well know, the big bang nucleosynthesis (BBN) provides very stringent constraints on the evolution of the early universe \cite{Cyburt:2015mya, Iocco:2008va, Steigman:2010zz, Kneller:2004jz, Steigman:2005uz, Steigman:2006yn, Pospelov:2010hj, Coc:2014oia} and has been used to test cosmological models (for example, see references \cite{Fritzsch:2012qc, Nesseris:2009jf, Tian:2015jjw, Dutta:2009jn, Boran:2013pux, Arik:2013sti, Ichimasa:2014fea, Clifton:2005xr, Coc:2006rt, Simha:2008zj, Lambiase:2012fv}). From equations (\ref{motfd1}) and (\ref{er}) and taking $0.85<s<1.15$ \cite{Steigman:2005uz}, we find the parameter $\alpha$ is constrained as $-0.141<\alpha<0.09$ or $1.000025<\alpha<1.000046$.

For a dust-dominated epoch, $w=0$, equation (\ref{integ}) reduces to
\begin{eqnarray}
\rho=\rho_0a^{\frac{6(\alpha-1)}{2-3\alpha}}.
\end{eqnarray}
We get the standard evolution of dust: $\rho=\rho_0a^{-3}$ for $\alpha=0$. If $\alpha>1$, in other words, the quantum fluctuations are large, we find that the expansion of the universe will be accelerated even in a dust-dominated period. However, it also can be expected that the effects of quantum fluctuations in the this epoch are very small. Taking $\alpha=0.01$, we get $\rho=\rho_0a^{-3-\frac{3}{197}}\simeq \rho_0a^{-3}$. Taking $\alpha=-0.01$, it is easy to have $\rho=\rho_0a^{-3+\frac{3}{203}}\simeq \rho_0a^{-3}$.

Now we consider an era dominated by dark energy (about $70\%$) and dark matter. The total EoS ($w=\Omega_{\rm D0}w_{\rm D0}$ where $\Omega_{\rm D0}$ and $w_{\rm D0}$ are respectively the present values of the dimensionless energy density parameter and the EoS of the dark energy) must be larger than that of dark energy which is close to $-1$ according to the Planck observations \cite{Ade:2015lrj}, the quantum fluctuations must be smaller than the classical metric $|\alpha |<1$. In order to obtain an accelerated phase, we also have to have $\alpha<\frac{3}{4}+\frac{1}{4w}$. We take $w=-0.7$ and $\alpha=0.01$ for approximate estimation. For these values of parameters $w$ and $\alpha$, we easily get $\rho\simeq \rho_0a^{-1782/1963}$ and $\frac{\ddot{a}}{a}=\frac{k^2}{6}\frac{536}{495}\rho>0$. Taking $w=-0.7$ and $\alpha=-0.01$, we have $\rho\simeq \rho_0a^{-606/679}$ and $\frac{\ddot{a}}{a}=\frac{k^2}{6}\frac{564}{505}\rho>0$.

For a dark energy dominated epoch, assuming a constant EoS, it is possible to obtain a decelerated expansion for $\frac{3}{4}+\frac{1}{4w}<\alpha<1$ due to the quantum fluctuations of metric. See, for example, taking $w=-9/10$ and $17/36<\alpha<1$, we get $\rho=\rho_0a^{\frac{6(\alpha-1)}{20-39\alpha}}$ and $\ddot{a}/a=-\frac{k^2}{6}\left[\frac{17-36\alpha}{10(1-\alpha)}\right]\rho<0$; taking $w=-11/10$ and $23/44<\alpha<1$, we get $\rho=\rho_0a^{\frac{6(\alpha-1)}{-20+41\alpha}}$ and $\ddot{a}/a=-\frac{k^2}{6}\left[\frac{23-44\alpha}{10(1-\alpha)}\right]\rho<0$, meaning that a decelerated expansion can be obtained in both cases.

\section{Conclusions and discussions}
We have considered a model of modified gravity from the nonperturbative quantization of a metric. We have derived the equations of the field and have applied it to the FLRW spacetime. We have investigated the effects of quantum fluctuations of metric on the very early universe, inflation, radiation-dominated, dust-dominated, and dark energy dominated epochs.
Due to the quantum fluctuations, it is possible to obtain both a bounce universe and a decelerated expansion in a dark energy dominated epoch for $\alpha=2/(3-w)$, $-1<w<-1/3$, and $\frac{3}{4}+\frac{1}{4w}<\alpha<2/3$. We have discussed the effects of quantum fluctuations on inflation parameters, such as slow-roll parameters, spectral index, and the spectrum of the primordial curvature perturbation. Values of $\alpha$ ($|\alpha |<1$ and $\alpha<$min$\{\frac{1}{2}, \frac{3}{4}+\frac{1}{4w}\}$) comparing the predictions of inflation can work to drive the current epoch of acceleration. If the quantum fluctuations are large, a accelerated expansion can be obtained even in a radiation-dominated or dust-dominated epoch (of cause, in fact, these two accelerated phase can not happen because the quantum part of the metric is small in these two era). We applied BBN to constrain the parameter and found it should satisfy $-0.141<\alpha<0.09$ or $1.000025<\alpha<1.000046$ to avoid conflicts with observations. Studies on the model proposed here, such as constraints from other astronomical observations or particle rate creation (see, for example, some related researches \cite{Pereira:2008at, Harko:2015pma, Erickcek:2013dea, Fabris:2014fda, Chakraborty:2014fia, Baranov:2015eha, Bertolami:2013uwl}), can be considered in the coming progresses.

Other cases, such as $\langle\delta \hat{g}^{\mu\nu}\rangle\propto R^{\mu\nu}$ or $G^{\mu\nu}$, are also worth discussing in future studies.

\begin{acknowledgments}
This study is supported in part by National Natural Science Foundation of China (Grant Nos. 11147028 and 11273010), Hebei Provincial Natural Science Foundation of China (Grant No. A2014201068), the Outstanding Youth Fund of Hebei University (No. 2012JQ02), the Open Project Program of State Key Laboratory of Theoretical Physics, Institute of Theoretical Physics, Chinese Academy of Sciences, China (No.Y4KF101CJ1), and the Midwest universities comprehensive strength promotion project.
\end{acknowledgments}
\bibliographystyle{elsarticle-num}
\bibliography{refq}

\end{document}